

\input harvmac

\overfullrule=0pt


\def\a{\alpha}

\def\b{\beta}

\def\g{\gamma}

\def\d{\delta}
\def\D{\Delta}
\def\e{\epsilon}

\def\th{\theta}

\def\m{\mu}
\def\n{\nu}

\def\l{\lambda}

\def\no{\noindent}

\def\bs{\bigskip}

\def\qq{\qquad}

\def\bl{\bigl}
\def\br{\bigr}


\def\IR{\relax{\rm I\kern-.18em R}}


\no hep-th/9205xxx \hfill USC-92/HEP-B2

\hfill                    May 1992

\bs\bs

 \centerline   {\bf CONFORMALLY EXACT METRIC AND DILATON}
\centerline  { {\bf IN STRING THEORY ON CURVED SPACETIME}
                   {\footnote{$^*$}
  {Research supported in part by DOE, under Grant No. DE-FG03-84ER-40168} } }

\vskip 0.75 true cm

\centerline {I. BARS and K. SFETSOS}

\bigskip

\centerline {Physics Department}
\centerline {University of Southern California}
\centerline {Los Angeles, CA 90089-0484, USA}


\vskip 1.00 true cm

\centerline{ABSTRACT}
\medskip

Using a Hamiltonian approach to gauged WZW models, we present a general method
for computing the conformally exact metric and dilaton, to all orders in the
$1/k$ expansion, for any bosonic, heterotic, or type-II superstring
model based on a coset $G/H$. We prove the following relations:
(i) For type-II superstrings the conformally exact metric and dilaton are
identical to those of the non-supersymmetric {\it semi-classical} bosonic
model except for an overall renormalization of the metric obtained by $k\to k-
g$. (ii) The exact expressions for the heterotic superstring are
derived from their exact bosonic string counterparts by shifting the central
extension $k\to 2k-h$ (but an overall factor  $(k-g)$ remains unshifted).
(iii) The combination $e^\Phi\sqrt{-G}$ is independent of $k$ and therefore
can be computed in lowest order perturbation theory as required by the correct
formulation of a conformally invariant path integral measure. The general
formalism is applied to the coset models $SO(d-1,2)_{-k}/SO(d-1,1)_{-k}$ that
are relevant for string theory on curved spacetime. Explicit expressions for
the conformally exact metric and dilaton for the cases $d=2,3,4$ are given. In
the semiclassical limit $(k\to \infty)$ our results agree with those obtained
with the Lagrangian method up to 1-loop in perturbation theory.

\bs\bs
\vfill
\eject


\newsec{Introduction}

During the past year there has been extensive investigations of
curved space-time string backgrounds generated by non-compact cosets $G/H$.
All models with space-time dimension $d\le 4$ require the non-compact current
algebra coset $SO(d-1,2)_{-k}/SO(d-1,1)_{-k}$  as part of, or as the full,
conformal field theory
\ref\BN{I. Bars and D. Nemeschansky, Nucl. Phys. {\bf B348} (1991) 89.}.
 The action is written in the form of a gauged WZW model
 \ref\WZW{E. Witten, Nucl. Phys. {\bf B223} (1983) 422.\semi
K. Bardakci, E. Rabinovici and B. Saering, Nucl. Phus. {\bf B301} (1988) 151.
\semi
K. Gawedzki and A. Kupiainen, Nucl. Phys. {\bf B320} (1989) 625. \semi
H. J. Schnitzer, Nucl. Phys. {\bf B324} (1989) 412.\semi
D. Karabali, Q-Han Park, H. J. Schnitzer and Z. Yang,
Phys. Lett. {\bf 216B} (1989) 307.\semi
D. Karabali and H. J. Schnitzer, Nucl. Phys. {\bf B329} (1990) 649.}.
For models involving more than four space-time coordinates there are other
possibilities which have been classified
\ref\Bf{I. Bars, ``Curved Space--Time Strings and Black Holes'',
in Proc. of {\it XX$^{th}$ Int. Conf. on Diff.
Geometrical Methods in Physics}, Eds. S. Catto and A. Rocha, Vol.2, p.695,
(World Scientific, 1992).\semi
P. Ginspang and F. Quevedo, ``Strings on Curved Space-Times:
Black Holes, Torsion, and Duality, LA-UR-92-640.},
but so far not investigated. The semi-classical analysis
\ref\WIT{E, Witten, Phys. Rev. {\bf D44} (1991) 314.}
for $k\to \infty$ has shown that
these are useful models for learning more about string and particle
propagation in gravitationaly singular spaces such as black holes and more
interesting singularities in various dimensions.
By now essentially all models
up to dimension four have been subjected to the semi-classical analysis
\ref\CRE{M. Crescimanno, Mod. Phys. Lett. {\bf A7} (1992) 489.}
\ref\HOHO{J. H. Horne and G. T. Horowitz, Nucl. Phys. {\bf B368} (1992) 444.}
\ref\BSf{I. Bars and K. Sfetsos, Mod. Phys. Lett. {\bf A7} (1992) 1091.}
\ref\BSs{I. Bars and K. Sfetsos, Phys. Lett. {\bf 277B} (1992) 269.}
\ref\FRA{E. S. Fradkin and V. Ya. Linetski, Phys. Lett. {\bf 277B} (1992) 73.}
\ref\HOR{P. Horava, Phys. Lett. {\bf 278B} (1992) 101.}
\ref\RAI{E. Raiten, ``Perturbation of a Stringy Black Hole'',
Fermilab-Pub 91-338-T.}
\ref\GER{D. Gerson, ``Exact Solutions of Four-Dimensional Black
Holes in String Theory'', TAUP-1937-91.}.
A cosmological interpretation has also been found
\ref\VAFA{A. Tseytlin and C. Vafa, Nucl. Phys. {\bf B372} (1992) 443.}
\ref\BSt{I. Bars and K. Sfetsos, ``Global Analysis of New Gravitational
Singularities in String and Particle Theories'',
USC-92/HEP-B1 (hep-th/9205037).}
\ref\Kounnas{C. Kounnas and D. L\"ust, ``Cosmological String Backgrounds
from Gauged WZW-models.}.
A group theoretical method for the global analysis of these semi-classical
geometries, including an explicit solution of the geodesics, has been
formulated and explicitly applied to some cases \BSt.

As in \BSs\ heterotic and type-II superstring actions can be constructed in
exactly 4 dimensions in the form of $N=1$ superconformal gauged WZW model. We
believe that a heterotic string model of this type, perhaps with some
variations, taken with a cosmological interpretation, provides the kind of
setting suitable for a discussion of the physics of the early Universe in the
context of string theory.

The principal method of semi-classical investigation followed ref. \WIT\ that
used a Lagrangian method. Quantum corrections which were necessary to obtain
the dilaton and satisfy the perturbative equations for conformal invariance
\ref\CAL{C. Callan, D. Friedan, E. Martinec and M. Perry,
Nucl. Phys. {\bf B262} (1985) 593.}
were limited to one loop. In practical terms one cannot carry out
the quantum computation of the sigma-model like theory to all orders
with this method.
However, the main interest in these models stem from the fact that
they are conformally exact current algebra theories, which are in principle
exactly solvable quantum theories. In order to take advantage of this fact it
is desirable to go back to the Hamiltonian method and use the algebraic
properties of the current algebra. The model can then be investigated via the
coset methods for non-compact current algebras
\ref\DB{L. Dixon, J. Lykken and M. Peskin, Nucl. Phys.
{\bf B325} (1989) 325.}
\ref\IB{I. Bars, Nucl. Phys. {\bf B334} (1990) 125.}.

In this paper we will show how to use the Hamiltonian approach to compute the
gravitational metric and dilaton backgrounds to all orders in the quantum
theory (all orders in the central extension $k$). These will then provide a
more accurate representation of the conformally exact vacuum configuration of
the string at the ``classical" level (i.e. no string loops). We have managed
to obtain these quantities for bosonic, type-II supersymmetric, and heterotic
string theories in $d\le 4$.
 \foot{The corresponding results are also given for a particle theory whose
WZW-like action was defined in \BSt. The particle theory can be thought of as
a string shrunk to a point which has no interactions with string excitations.
For this case the semi-classical result is actually exact.}

The main idea is the following: the conformally exact Hamitonian is the sum of
left and right Virasoro generators $L^L_0+L_0^R$ that may be written purely in
terms of Casimir operators of $G$ and $H$ at the tachyon level. The exact
dependence on the central extension $k$ is included in this form. If we
investigate the exact quantum eigenstates in configuration space, then the
Casimir operators become Laplacians constructed as differential operators in
group parameter space ($dim G$). If the state $\psi$ is a singlet under the
gauge group $H$ (acting simultaneously on left and right movers), then gauge
invariance requires that it is a function of singlet combinations of group
parameters. There are exactly $dim(G/H)$ such invariants which we choose as
our string coordinates $X^a$. We have recently shown \BSt\ that these
invariants provide a global description of the geometry. In
this way we can write the conformally exact Hamiltonian $L_0^L+L_0^R$ as a
differential operator in the global curved space-time manifold involving only
the string coordinates $X^a$. By comparing to the expected general form
$(L_0^L+L_0^R)\psi={-1\over e^\Phi\sqrt{-G}}\partial_a(e^\Phi\sqrt{-
G}G^{ab}\partial_b\psi)$ for the singlet $\psi$, we read off the exact global
metric and dilaton.

We have applied this program to the general bosonic, heterotic and type-II
superstrings and derived relationships among the exact quantities of these
theories as announced in the abstract of this paper. For the specific cosets of
interest $SO(d-1,2)/SO(d-1,1)$ explicit expressions are given below.
The large $k$ limit of our results agree with the previous semi-classical
computations. In the special case of two dimensions we also agree with another
previous derivation of the exact metric and dilaton for the $SO(2,1)/SO(1,1)$
bosonic string
 \ref\DVV{R. Dijgraaf, E. Verlinde, H. Verlinde, Nucl. Phys. {\bf B371} (1992)
269.}.


\newsec { Algebraic formalism for computing the exact metric and dilaton }

Let us consider a bosonic string theory for closed strings in $d$ curved
space-time dimensions, based on a sigma model conformal field theory
with string coordinates $X^a,\ a=0,1\cdots d-1 $. The space-time metric and
dilaton fields are $G_{ab}(X)$ and $\Phi (X)$ respectively. We begin with the
effective action for the tachyon field $T(X)$ in $d$ space-time dimensions.
The most general form of this effective action is

\eqn\action{\eqalign{&S[T]=\int d^d X \sqrt{-G} e^{\Phi}
\bl(G^{ab}\partial_a T \partial_b T -V(T)\br)\cr
&V(T)=2T^2+{\cal O}(T^3)\ ,\cr} }

\no
where $V(T)$ is the tachyon potential whose precise form is not
necessary for the analysis that follows. From the point of view of
conformal field theory
the tachyon is completely defined through the action of the zero
modes $L^L_0$ and $L^R_0$ of the stress tensors
for the left and the right movers. Therefore \action\ must be equivalent
to the following action

\eqn\caction{S[T]=\int d^d X \bl(T(L^L_0 +L^R_0)T -V(T)\br)\ .}

\no
Comparison of \action\ with \caction\ determines the form of
$L^L_0 +L^R_0$ as a differential operator in configuration space

\eqn\lzero{(L^L_0 +L^R_0)T=-{1\over e^{\Phi}\sqrt{-G}}
\partial_a (G^{ab}e^{\Phi}\sqrt{-G}\ \partial_b T)\ .}

Now let us consider the sigma model like action which results from an exact
conformal theory based on the gauged WZW action. Using the equivalent current
algebra coset model $G/H$ we can write $L^L_0$ in terms of the quadratic
Casimir operators $\D^L_G$ and $\D^L_H$ for the group and the subgroup, as
follows

\eqn\lzer{\eqalign{&L^L_0 T=\bl({\D^L_G\over k-g}-{\D^L_H\over k-h}\br)T\cr
&\D^L_G \equiv Tr(J^L_G)^2, \qq \D^L_H \equiv Tr(J^L_H)^2\ ,\cr} }

\no
where $J^L_G$ and $J^L_H$ are {\it antihermitian} group and subgroup generators
obeying the appropriate Lie algebras, and $g$, $h$ are the Coxeter numbers for
the group and the subgroup. For the cases of interest in this paper $g=d-1$,
$h=d-2$ for $d\ge 3$, and $g=2,\ h=0$ for $d=2$.
 \foot{For the particle theory of footnote 1 the Hamiltonian contains no
Coxeter numbers since the higher string excitations are absent. Then $L^L_0=
(\D^L_G-\D^L_H)/k$.}
 An expression similar to \lzer\  can also be written for $L^R_0$. As shown
below, we construct the generators $J^L_G,J_G^R,J^L_H,J_H^R$ as first order
differential operators acting on group parameter space. Then the Casimir
operators $\D^L_G$, $\D^L_H$,$\D^R_G$, $\D^R_H$, contain single and double
derivatives with respect to all $dimG$ parameters in $G$.
 \foot{Since we have defined our Casimir operators as the square of
antihermitian generators we differ by a minus sign from usual conventions. For
example for $SU(2)$ we would get the eigenvalues $\D_G=-j(j+1)$ instead of
$+j(j+1)$.}

Gauging the subgroup $H$ means that we have to impose the following gauge
invariance conditions on the tachyon $T$
 \foot{If $H$ contains an abelian $U(1)$ or $\IR$ factor there is the
alternative of imposing the axial gauging condition $(J^L_H-J^R_H)\ T=0$ for
the currents associated with the abelian factor. For an application see
\ref\SF{K. Sfetsos, in preparation.}.}

\eqn\cond{(J^L_H +J^R_H)\ T=0\ .}

\no
The number of conditions is $dimH$ and therefore $T$ can depend only on
$d=dim(G/H)$ parameters, $X^a$ (string coordinates), which are $H$-invariants.
The fact that there are exactly $dim(G/H)$ such independent invariants is not
immediately obvious but it should become apparent to the reader by considering
a few specific examples. As discussed in \BSt\ these are in fact the
coordinates that globally describe the sigma model geometry. Consequently,
using the chain rule, we reduce the derivatives in \lzer\ to only derivatives
with respect to the $d$ string coordinates $X^a$. Moreover, using the
fact that $\D^L_G=\D^R_G$ for any group
 \foot{This follows from $J^R_G = -g^{-1}J^L_G\ g +{1\over D}
Tr(g^{-1}J^L_G\ g)$, where $D$ is the dimension of the matrix $g$. The
second term is present because $g$ and $J_G^L$ do not commute as quantum
operators and $J_G^R$ has to be traceless. But despite its
presence the relation $\D_G^L=\D_G^R$ is derived
from it.},
 together with the fact that the gauge invariance condition \cond\ leads to
$(\D^L_H -\D^R_H)T=0$ (see (2.14) below), we ensure the physical condition for
closed bosonic strings $(L^L_0 -L^R_0)T=0$. Then using \lzero\ and \lzer\ one
can deduce uniquely the expression for the inverse metric $G^{ab}$ by
comparing  the coefficients of the double derivatives $\partial_a\partial_b
T$. Comparison of the single derivative terms $\partial_aT$ will give a system
of $d$ coupled linear partial differential equations, whose solution
determines the dilaton field $\Phi$.

The general $k$ dependence of the exact expressions takes a particular form
that can be seen as follows. In the large $k$ limit \lzer\ becomes
proportional to ${1\over k}(\D_G^L-\D_H^L)$ from which we can read off the
semi-classical metric and dilaton according to the above procedure. Therefore,
we may rewrite \lzer\ in the following form

\eqn\lze{ L_0^LT={1\over k-g}(\D^L_{G/H}+{g-h\over k-h}\D^L_H)T}
where $\D^L_{G/H}=\D^L_G-\D^L_H$. Then it is evident that, except for the
overall factor $(k-g)$, all dependence on $k$ has the form ${g-h\over k-h}$.
This applies to the bosonic string. For the heterotic and type-II superstrings
the $k$ dependence can be derived by the same technique as will be seen below.
It is evident that for the particle theory of footnotes 1,2 there are no such
corrections to the semi-classical result.

Let us specialize to the  coset models $SO(d-1,2)_{-k}/SO(d-1,1)_{-k}$ with
$d=2,3,4$ since these are the ones of interest for a theory in four dimensions.
We want to find the currents appropriate for right or left transformations
of the group elements of $SO(d-1,2)$ in a $SO(d-1,1)$ basis. It is convenient
to parametrize the group element of $SO(d-1,2)$ as the product $g=h t$, where
$h\in SO(d-1,1)$ and $t\in SO(d-1,2)/SO(d-1,1)$. The $h$, $t$ are given by
 \foot{We follow the notation of \BSt. As explained there, to compare with
\BSf\BSs\ where another vector $X^{\m}$ was used, one should set
$X^{\m}=2x^{\m}/(x^2-1)$. These Lorentz vectors should not be confused with
the Lorentz invariant string coordinates $X^a$ even though they have the same
dimension $d=dim(G/H)$.}

\eqn\htt { h=\left ( \matrix {1 & 0 \cr 0 & h_\mu^{\ \nu} \cr } \right ) ,
\qquad \qquad t=\left (\matrix {b  & (b+1) x^\nu \cr
-(b+1) x_\mu  & (\eta_\mu^{\ \nu} -(b+1) x_\mu x^\nu) \cr } \right ).}

\no
Furthermore $h$ can be written in the form $h_{\m}{}^{\n}= [(1+a)(1-a)^{-
1}]_{\m}{}^{\n}$, with $a_{\mu\nu}=-a_{\nu\mu}$ when both indices are lowered.
To insure that $t$ is a $SO(d-1,2)$ group element we take $b={1-x^2\over
{1+x^2}}$. By considering the infinitesimal left transformations $\d_L g=\e_L
g$ we can read off the form of the generators

\eqn\leftgen{\eqalign{&J^L_{\m\n}={1\over 2}(1+a)_{\m\a}(1+a)_{\n\b}
{\partial\over \partial a_{\a\b}}\cr
 &J^L_{\m}=-{1\over 2}(1+x^2)\bl({1+a\over 1-a}\br)_{\m}{}^{\n}
{\partial \over
\partial x^{\n}} +{1\over 2}(1+a)_{\m\a}(1+a)_{\b\g}x^{\g}
{\partial\over \partial a_{\a\b}}\ .}  }

\no
It can be shown that the above generators obey the commutation rules of the
$SO(d-1,2)$ algebra. Namely

\eqn\leftcom{\eqalign{&\bl[J^L_{\m\n},J^L_{\a\b}\br]=
J^L_{\m\a}\eta_{\n\b}-J^L_{\n\a}\eta_{\m\b}
              +J^L_{\n\b}\eta_{\m\a}-J^L_{\m\b}\eta_{\n\a}\cr
&\bl[J^L_{\m\n},J^L_{\a}\br]=\eta_{\m\a}J^L_{\n}-\eta_{\n\a}J^L_{\m}\cr
&\bl[J^L_{\m},J^L_{\n}\br]=J^L_{\m\n}\ .} }

\no
If we consider instead, the infinitesimal right transformations
$\d_R g=g \e_R$ we find the following expressions

\eqn\rightgen{\eqalign{&J^R_{\m\n}=
-{1\over 2}(1-a)_{\m\a}(1-a)_{\n\b}
{\partial\over\partial a_{\a\b}}-x_{[\m}{\partial \over \partial x^{\n]}}\cr
&J^R_{\m}={1\over 2}(x^2-1){\partial\over \partial x^{\m}}-x_{\m}x^{\n}
{\partial \over \partial x^{\n}}
-{1\over 2}(1-a)_{\m\a}(1-a)_{\g\b}x^{\g}
{\partial\over \partial a_{\a\b}}\ .}}

\no
These currents obey the same commutation rules as in \leftcom\ and moreover
commute with the left currents $[J^L,J^R]=0$. Now we construct the quadratic
Casimir associated with the left and right currents. We find

\eqn\casimir{\eqalign{\D^L_G&=
{1\over 2}(J^L)_{\m\n}(J^L)^{\m\n}+(J^L)_{\m}(J^L)^{\m}\cr
 &={1\over 4}(1+x^2)^2 {\partial^2\over \partial x^{\m}x_{\m}}
-{d-2\over 2}(1+x^2)x_{\m}{\partial\over \partial x_{\m}}\cr
&+{1\over 4}(1+a)_{\m\g}x^{\g}(1+a)_{\a\d}x^{\d}(1-a^2)_{\n\b}
{\partial^2\over \partial a_{\m\n}\partial a_{\a\b}}\cr
&+{1\over 2}(1+a)_{\m\a}x^{\a}(1-a^2)_{\n\b}x^{\b}
{\partial\over \partial a_{\m\n}}\cr
&+{1\over 2}(1+x^2)(1+a)_{\m\a}x^{\a}(1+a)_{\n\b}
{\partial^2\over \partial a_{\m\n}\partial x_{\b}}\cr
&+{1\over 8}(1-a^2)_{\m\a}(1-a^2)_{\n\b}
{\partial^2\over \partial a_{\m\n}\partial a_{\a\b}}
+{1\over 4}(a-a^3)_{\m\n}{\partial\over \partial a_{\m\n}}\ .} }

\no
The expression for $\D^R_G$ is identical for reasons explained in footnote 5.
The quadratic Casimir operators corresponding to the subgroup $H=SO(d-1,1)$
are

\eqn\Hcasimirleft{\eqalign{\D^L_H&={1\over 2}(J^L)_{\m\n}(J^L)^{\m\n}\cr
&={1\over 8}(1-a^2)_{\m\a}(1-a^2)_{\n\b}
{\partial^2\over \partial a_{\m\n}\partial a_{\a\b}}
+{1\over 4}(a-a^3)_{\m\n}{\partial\over \partial a_{\m\n}}\ .\cr} }

\no
and

\eqn\Hcasimirright{\eqalign{\D^R_H&={1\over 2}(J^R)_{\m\n}(J^R)^{\m\n}\cr
&={1\over 8}(1-a^2)_{\m\a}(1-a^2)_{\n\b}
{\partial^2\over \partial a_{\m\n}\partial a_{\a\b}}
+{1\over 4}(a-a^3)_{\m\n}{\partial\over \partial a_{\m\n}}\cr
&+(x^2\eta_{\m\n}-x_{\m}x_{\n})
{\partial^2\over \partial x_{\m}\partial x_{\n}}
-(d-1)x_{\m}{\partial\over \partial x_{\m}}
+(1+a)_{\m\a}x^{\a}(1+a)_{\n\b}{\partial^2\over \partial a_{\m\n}\partial
x_{\b}}\ .}  }

\no
The two expressions differ by the last line in \Hcasimirright\ which
equals to

\eqn\diff{\D^R_H-\D^L_H={1\over 2}\bl((J^R)^{\m\n}-(J^L)^{\m\n}\br)
\bl((J^R)_{\m\n}+(J^L)_{\m\n}\br)\ .}

\no
Using the expressions \leftgen\ \rightgen\ the gauge invariance conditions
\cond\ take the form

\eqn\gauge{\bl(a_{[\m\l} {\partial\over \partial a_{\l}{}^{\n]}}
-x_{[\m} {\partial\over \partial x^{\n]}}\br)\ T=0\ ,}

\no
where the brackets indicate antisymmetrization of the $\mu,\nu$ indices. This
form is recognized as the global Lorentz generator and it requires that $T$ be
constructed only from Lorentz invariants that can be formed from
$x_\mu$ and $a_{\mu\nu}$. Next we specialize to the cases $d=3$ and $d=4$. The
$d=2$ case corresponding to the $2d$ black hole is discussed in the Appendix.



\newsec { 3d Bosonic String}

The $3d$ model based on the coset model $SO(2,2)/S(2,1)$ was discussed
semi-classically from the gauged WZW model Lagrangian point of view in
\CRE\BSf\FRA. This model may be viewed as the $3d$ submanifold of a four
dimensional model which is constructed by adjoining a factor of $U(1)$ or
$\IR$ to the coset. The global structure of the 3d manifold was analyzed in
\BSt\ by finding the global coordinates and examining the particle
trajectories. In particular, it was found that the space consists of two
topologically distinct sectors. There is a curvature singularity with the
topology of ``pinched double trousers'' in one sector and that of a ``double
saddle'' in the other. In this case the antisymmetric matrix $a_{\m\n}$ has
three parameters, therefore it is possible to reparametrize it in terms of a
three dimensional vector $y^{\m}$, as $a_{\m\n}=\e_{\m\n\l}y^{\l}$. Then the
gauge condition \gauge\ takes the form

\eqn\trigauge{\bl(y_{[\m}{\partial\over \partial y^{\n]}}
+x_{[\m}{\partial\over \partial x^{\n]}}\br)\ T=0\ .}



\no
The constraint \trigauge\ requires that $T$ depend only on the three Lorentz
invariants $x^2$, $y^2$, $x\cdot y$, or their combinations. In fact, in
order to make correspondence with previous results we choose the same
invariants as in \BSt\ with $T(v,u,b)$ where

\eqn\tristates{ b={1-x^2\over 1+x^2}, \qq v={2\over 1+y^2},
\qq u=-2{(x\cdot y)^2 \over x^2 (1+y^2) }\ .  }

\no
Using the chain rule we transform the derivatives with respect
to the vectors $x^{\m}$ and $y^{\m}={1\over 2}\e^{\m\n\l}a_{\n\l}$
in \casimir\ \Hcasimirleft\ to
derivatives with respect to the $H$--invariants $v$, $u$, $b$,
e.g. ${\partial \over \partial x_{\m}}T=
2v (x\cdot y) /x^4 ((x\cdot y)x^{\m}-x^2 y^{\m}){\partial \over \partial u}T
-(b+1)^2 x^{\m} {\partial \over \partial b}T $, etc. Finally, with the dot
products in the Laplacians, $L_0$ is written only in terms of $(v,u,b)$ when
acting on $T$. Then comparison of the double derivatives terms in \lzero\ and
\lzer\ gives the non-zero elements of the inverse of the metric (we omit an
overall factor of $1\over 2(k-2)$ which will be restored later in $(3.5)$)

\eqn\invetrimetric{\eqalign{&G^{bb}=4(b^2-1)\cr
&G^{vv}=-4{b-1\over b+1}v(v-u-2)+{4\over k-1}v(v-2)\cr
&G^{uu}=4{b+1\over b-1}u(v-u-2)+{4\over k-1}u(u+2)\cr
&G^{vu}={4\over k-1}vu\ ,\cr}  }

\no
and comparison of the single derivative terms yields a system of linear
partial differential equations which determine the dilaton

\eqn\tridiff{\eqalign{
{\partial\over \partial b}ln\bl(\sqrt{-G}e^{\Phi}\br)&=
{b\over b^2-1}\cr
{\partial\over \partial v}\bl(G^{vv}\sqrt{-G}e^{\Phi}\br)+
{\partial\over \partial u}\bl(G^{uv}\sqrt{-G}e^{\Phi}\br)&=
 2 \sqrt{-G}e^{\Phi}\bl[
{b-1\over b+1}(u+2-3v) +{4v-2\over k-1}\br]\cr
{\partial\over \partial u}\bl(G^{uu}\sqrt{-G}e^{\Phi}\br)+
{\partial\over \partial v}\bl(G^{vu}\sqrt{-G}e^{\Phi}\br)&=
 2 \sqrt{-G}e^{\Phi} \bl[{b+1\over b-1} (v-2-3u)
+{4u+2)\over k-1}\br]\ .\cr}  }

\no
Note that without a dilaton these equations have no solutions. Therefore, even
without the hindsight of general string arguments, a dilaton must be introduced
in our approach in order to have a solution to these equations.
If we invert the inverse metric we get for the line element the following
expression (we also restore the overall factor $2(k-2)$)

\eqn\trimetric{ds^2=2(k-2)\bl(G_{bb} db^2 +G_{vv} dv^2 +G_{uu} du^2
+2G_{vu} dvdu\br)\ . }

\no
with
\eqn\tridef{\eqalign{&G_{bb}={1\over 4(b^2-1)}\cr
&G_{vv}=-{\b(v,u,b)\over 4v(v-u-2)}
\bl({b+1\over b-1} + {1\over k-1} {u+2\over v-u-2}\br)\cr
&G_{uu}={\b(v,u,b)\over 4u(v-u-2)}
\bl({b-1\over b+1} - {1\over k-1} {v-2\over v-u-2}\br)\cr
&G_{vu}={1\over 4(k-1)} {\b(v,u,b)\over (v-u-2)^2}\ ,\cr}  }

\no
and where the function $\b(v,u,b)$ is defined as follows

\eqn\defb{\b^{-1}(v,u,b)=1+{1\over k-1} {1\over v-u-2}
\bl({b-1\over b+1} (u+2) -{b+1 \over b-1} (v-2) -{2\over k-1}\br)\ . }

It remains to solve the system of differential equations \tridiff.
Although the solution to those equations is straightforward, it is
illuminating to guess the solution by recalling some of the results of
\BSf\BSs. There it was found that conformal invariance required a path
integral measure for gauged WZW models that includes, in addition to the
Haar measure for the group $G$, an extra purely group theoretical factor
which must be H--invariant.
 \foot{The appropriate factor can be obtained in a patch of the manifold by
choosing a unitary gauge. It must be designed to cancel the Haar
measure and the Faddeev-Popov determinant so that the dilaton is associated
only with the determinant that emerges from the integration of the gauge
fields. This dilaton then solves the one-loop conformal invariance
conditions. The extra factor is extended from the patch in the unitary gauge
to the $H-$invariant global manifold by the methods of \BSt.}
 Moreover, it was observed that this factor was equal to $1/(e^{\Phi} \sqrt{-
G})$. This result was true in the semi-classical limit $k\to \infty$. Because
of the group theoretic nature of the derivation it was conjectured that the
combination $e^{\Phi} \sqrt{-G}$ is $k$--independent and equal to the 1--loop
result, although individually the metric and the dilaton can receive $1/k$
corrections. If the conjecture is not correct it implies that the path
integral measure for the gauged WZW model is not defined once and for all at
the outset, before one begins a perturbative analysis. In the notation of \BSt
the one loop semi-classical result is

\eqn\semic{e^{\Phi} \sqrt{-G} {\big \vert}_{k\to \infty}
= \bl({b^2-1\over vu}\br)^{1\over 2}\ . }

\no
One can now check that this expression indeed satisfies
the system of differential equations \tridiff\ for all values of $k$.
Therefore, the conjecture is correct as we expected on the basis of a well
defined path integral quantization of the model. So we have proven the theorem

\eqn\theorem{ e^\Phi\sqrt{-G}=e^\Phi\sqrt{-G}{\big \vert}_{k \to\infty}
                \qquad {\rm for\ all\ } k . }
This is also true in all string and superstring models we consider in
the present paper. We are convinced that this is a general feature of gauged
WZW models. After calculating $\sqrt{-G}$ from \invetrimetric\ or \tridef\ the
result for the dilaton is the following

\eqn\tridilaton{{\Phi}=\ln \bl({ (b^2-1) (v-u-2)\over \sqrt{\b(v,u,b)}}\br)
+ \Phi_0\ , }

\no
where $\Phi_0$ is the constant of integration.
In the limit $k\to \infty$, $\b \to 1$ both the metric and the dilaton
tend to their semi-classical expressions of \BSt.

One might ask the question: how does the finite value of $k$ modify the
manifold? In figures 1.ab, 2.ab, and 3.ab the allowed
regions in the $v-u$ plane are indicated at fixed values of $b$. As in \BSt,
the three signs inside the parentheses in the various regions are the signs of
the coefficients of $dv^2$, $du^2$ and $db^2$ in the semi-classical metric. A
minus (plus) sign corresponds to a time (space) coordinate, thus indicating
the signature of the region. The regions with one time coordinate correspond
to the $SO(2,2)/SO(2,1)$ coset. The remaining regions correspond to the
analytic continuations $SO(3,1)/SO(3)$ $(+\ +\ +)$, $SO(4)/SO(3,3)$ $(-\ -\ -
)$, and $SO(3,1)/SO(2,1)$ (one plus). Thus, by specializing to each one of
these regions our exact metric and dilaton describe those cosets as well. The
$45^o$ line $u=v-2$ is a curvature singularity in the semi-classical limit,
the other two being at $b=\pm 1$. The way the $b$-fixed planes are sliced up
by the lines at $u=0,\ v=0,\ v-u-2=0$ into regions of various signatures is a
purely group theoretical result about the coset manifolds that are listed
above. That is, the coset manifold $SO(2,2)/SO(2,1)$ lives in the ranges of
$(v,u,b)$ parameters indicated in the figures, independently than any metric
(similarly for the other manifolds). As it turns out, the full region
coincides with the properties of the semi-classical metric. However, quantum
corrections may require additional constraints on the acceptable regions in
order to maintain the signature. This is indeed what happens, and how the
$k$-dependence of the exact metric shows up. The second line, with varying
slope (which depends on $b$) is a singularity of the function $\b(v,u,b)$.
For the case of the coset $SO(2,2)/SO(2,1)$ and $k>2$\
 \foot{ The signature depends crucially on the sign of $(k-2)$ as seen from
$(3.5)$. Demanding $c=26$ for the bosonic string gives $k\simeq 2.48$ or
$k\simeq 0.91$, and $c=15$ for the superstring gives $k=20/7$. It is believed
that a consistent quantum theory requires $k>2$  \IB. Nevertheless one could
perform a similar analysis even when $k<1$.}
 one must demand that $\b(v,u,b) >0$ so that the exact metric has the correct
signature, as seen from the determinant of the metric (one time coordinate
requires $detG<0$). This leads to further restrictions for the allowed
regions. The result is shown as the shaded areas in the figures: they have
switched signature due to the quantum corrections. Therefore, although they
were previously allowed, they are now off limits since a classical geodesic in
the $SO(2,2)/SO(2,1)$ white regions cannot enter the shaded areas. Therefore
quantum amplitudes are expected to decay off and tunnel in these regions. This
implies that quantum effects have created a screening of the classical
singularity, although not everywhere. In particular, using the figures of
\BSt\ , we indicate the 3-dimensional view of the screening that occurs. The
completely screened singularities are: (i) the central blob of the
double trouser singularity in region II, (ii) the pinched region and the
$b\sim \pm 1$ neigborhoods on the inside of the trousers, and (iii) the entire
sheets forming the saddle singularities in regions I' and III'. The unscreened
regions are: (i) Part of the inside of the trousers in regions
I and III when $|b|>k/(k-2)$, (ii) the farther out parts of the outside of the
trousers in region II, and (iii) the region II' between the saddles.


\newsec {4d Bosonic String}

The $4d$ model based on the coset $SO(3,2)/SO(3,1)$ was analyzed in \BSs\
where expressions for the perturbative metric and dilaton were given in some
patches of the manifold. We will see that the method we have been
following in the present paper leads to the discovery of the global
coordinates as well. The four invariants one can construct and which satisfy
the gauge condition \gauge\ are

\eqn\testates{x^2\ ,\qq z_1={1\over 4}Tr(a^2)\ ,\qq
z_2={1\over 4}Tr(a^* a)\ , \qq z_3=xa^2x/x^2 \ ,}

\no
where $a^*_{\m\n}={1\over 2}\e_{\m\n\a\b}a^{\a\b}$ is the dual of $a_{\m\n}$.
However, the metric written in these coordinates is non-diagonal and
very complicated. Instead we use a different set of four invariants
$b$, $u$, $v$, $w$ for which the semi-classical metric is diagonal,

\eqn\relat{\eqalign{&b={1-x^2 \over 1+x^2}\ , \qq
u={1+z_2^2+2(z_1 -z_3) \over 1-2 z_1-z_2^2}\cr
&v={1+z_1 +\sqrt{z_1^2 +z_2^2} \over 1-z_1 -\sqrt{z_1^2 +z_2^2}}\ ,\qq
w={1+z_1 -\sqrt{z_1^2 +z_2^2} \over 1-z_1 +\sqrt{z_1^2 +z_2^2}}\ .} }

\no
To find the ranges in which the above global coordinates take their values we
consider a Lorentz frame that can cover all possibilities without loss of
generality. First we notice that by Lorentz transformations the antisymmetric
matrix $a_{\m\n}$ can always be transformed, as in \BSs, to a block diagonal
matrix, with the non-zero elements

\eqn\blo{a_{01}=\tanh t\ {\rm or}\ \coth t\ ,\qq a_{23}=\tan \phi\ .}

\no
Then using \relat\ one can deduce the form of the global variables:
$v=\pm \cosh 2t$, $w=\cos 2\phi$, and $u={1\over x^2}
\bl(w(x_0^2-x_1^2)-v(x_2^2 +x_3^2)\br)$. Therefore the string variables
can take values in the following regions

\eqn\range{\eqalign{
&{\rm region-I} :\qquad  b^2>1\ ,\qq  w<u<v\ \ \ {\rm or}\ \ \ v<u<w\cr
&{\rm region-II} :\qquad  b^2<1\ ,\qq  -\infty<u<w<v\ \ \ {\rm or}\ \ \
v<w<u<+\infty\ .\cr}  }

\no
In both regions $v^2>1$ and $-1<w<1$.
Then by considering states of the type $T=T(b,u,v,w)$ and following
a procedure similar to the $3d$ case we find the line element

\eqn\temetric{\eqalign{ds^2=2(k-3)
&\bl(G_{bb} db^2 +G_{uu} du^2 +G_{vv} dv^2 +G_{ww} dw^2\cr
&+2 G_{uv} du dv +2 G_{uw} du dw +2 G_{vw} dv dw\bl)\ ,\cr} }

\no
where
\eqn\tedef{\eqalign{&G_{bb}={1\over 4(b^2-1)}\cr
&G_{uu}={\b(b,u,v,w)\over 4(u-w)(v-u)}\biggl({b-1\over b+1}-{1\over k-2}
{(v-w)^2\over {(v-u)(u-w)}} (1-{1\over k-2} {b+1\over b-1})\biggr)\cr
&G_{vv}=-{(v-w)\ \b(b,u,v,w)
\over 4(v^2-1)(v-u)}\biggl({b+1\over b-1}-{1\over k-2}
{1\over (v-u)(u-w)}\bl[1-u^2 \cr
 &\qq +({b+1\over b-1})^2 (v-u)(v-w)
   +{1\over k-2}
{b+1\over b-1} {(1+v^2)(u+w)-2v(1+uw)\over v-w}\br]\biggr)\cr
&G_{ww}={(v-w)\ \b(b,u,v,w)\over 4(1-w^2)(u-w)}
 \biggl({b+1\over b-1}-{1\over k-2}
{1\over (v-u)(u-w)}\bl[1-u^2 \cr
 & \qq +({b+1\over b-1})^2 (u-w)(v-w)
    -{1\over k-2}
{b+1\over b-1} {(1+w^2)(u+v)-2w(1+uv)\over v-w}\br]\biggr)\cr
&G_{uv}={\b(b,u,v,w)\over 4(k-2)(v-u)^2}\bl(1-{1\over k-2} {b+1\over b-1}
{v-w\over u-w}\br)\cr
&G_{uw}={\b(b,u,v,w)\over 4(k-2)(u-w)^2}\bl(1-{1\over k-2} {b+1\over b-1}
{v-w\over v-u}\br)\cr
&G_{vw}={1\over (k-2)^2} {b+1\over b-1} {\b(b,u,v,w)\over
4(v-u)(u-w)}\ ,\cr} }

\no
where the function $\b(b,u,v,w)$ is defined by

\eqn\defbb{\eqalign{\b^{-1}&=1+{1\over k-2} {(v-w)^2
\over (v-u)(w-u)} \biggl({b+1\over b-1}+{b-1\over b+1} {1-u^2\over (v-w)^2}\cr
&+{1\over k-2}\bl({vw+u(v+w)-3 \over (v-w)^2} -({b+1\over b-1})^2\br)\biggr)
+{2\over (k-2)^3} {b+1\over b-1} {vw-1 \over (v-u)(u-w)}\ .\cr} }

\no
The dilaton field is

\eqn\tedilaton{\Phi=\ln \bl({(b^2-1)(b-1)(v-u)(w-u)\over
\sqrt{\b(b,u,v,w)}}\br) +\Phi_0\ . }

\no
It would be instructive to write the expression for the metric in the
semi-classical limit
and verify that the range for
the string parameters \range\ is such that there is only one time
coordinate in every region. From \tedef\ we obtain

\eqn\semetr{\eqalign{{ds^2\over 2(k-2)} {\big \vert}_{k\to \infty}=
&{db^2\over 4(b^2-1)}+{b-1\over b+1} {du^2\over 4(v-u)(u-w)}\cr
&+{b+1\over b-1}(v-w)\bl({dw^2\over 4(1-w^2)(u-w)}
-{dv^2\over 4(v^2-1)(v-u)}\br)\ .\cr} }

\no
This metric was derived in \BSs\ in region--I.
It can be easily seen that in  region--I of \range\ ``time"$=v$
whereas in region--II ``time"$=b$.
The presence of a finite $k$ modifies the manifold in a similar way to
the $3d$ case, as considered in the previous section.

Again we can check that $e^\Phi\sqrt{-G}$ is independent of $k$, which proves
the theorem for $d=4$.


\newsec{ Type--II and Heterotic superstrings }

In this section we consider superconformal extensions of the bosonic
models we have been considering in the previous sections.
The general type--II and heterotic superstring model in curved
space-time was defined in \BSs\  in the form of supersymmetric $N=1$ gauged
WZW model. This corresponds to a Kazama-Suzuki model with a non-compact group
$SO(d-1,2)$ \BN\ and therefore can be analyzed with current algebra techniques
\DB\IB . The $4d$ case  was worked out explicitly, to leading order in
perturbation theory, using the Lagrangian method. As we shall see, the exact
metric and dilaton for these superstring models are closely related to the
corresponding expressions for the bosonic strings. It was pointed out in
footnote 5 that for any WZW model $\D^L_G=\D^R_G \equiv \D_G$. Restricting to
H--invariant tachyon states, as we were instructed to do by \cond, gives
another condition $\D^L_H T=\D^R_H T \equiv \D_H T$, or equivalently
$\D^L_{G/H} T=\D^R_{G/H} T \equiv \D_{G/H} T$. For the bosonic string we saw
that these remarks led to the $k$ dependence exhibited in \lze . Now we turn
to the superstrings.

For the type--II superstring the coset model with $N=1$ superconformal
symmetry ($N=2$ if $G/H$ is k\"ahlerian) is described by \BN

\eqn\super{ {SO(d-1,2)_{-k} \otimes SO(d-1,1)_1 \over SO(d-1,1)_{-k+g-h}}\ , }

\no
for both the left and the right movers. For the tachyon the fermionic factor
$SO(d-1,1)_1$ makes no contribution. However, the shifting of
the level in the denominator in \super\ , i.e. $(-k+g-h)=-k+1$ instead of $-k$
for the bosonic case, has a profound effect and we obtain the quantum
Hamiltonian with the exact dependence on $k$

\eqn\type{\eqalign{(L^L_0 +L^R_0)\ T_{II}&=\bl({\D^L_G \over k-g}
-{\D^L_H \over k-g} +{\D^R_G \over k-g}
-{\D^R_H \over k-g}\br)\ T_{II}\cr
&={2\over k-g}\D_{G/H}\ T_{II}\ .\cr} }

\no
Except for an overall renormalization of $k\to k-g$, this is
exactly the expression in the semi-classical limit.
Therefore, almost trivially we have
proven a theorem: For any type-II superstring based on a Kazama-Suzuki coset,
as in \super\ , the exact metric and dilaton are  given by the 1--loop
perturbative result except for the overall normalization of the metric!! In
the special case of $d=2$ available field theoretic perturbative computeations
verify this result up to 5 loops
 \ref\TSEY{A.A. Tsetylin, Phys. Lett. {\bf 268B} (1991) 175.
\semi I. Jack, D. R. T. Jones and J. Panvel, ``Exact Bosonic and
Supersymmetric String Black Hole Solutions", LTH-277.}.

For a heterotic superstring only the left sector is supersymmetric whereas
the right sector is not. Therefore, for the tachyon we can write

\eqn\hete{\eqalign{(L^L_0 +L^R_0)\ T_{het}&=\bl({\D^L_G \over k-g}
-{\D^L_H \over k-g} +{\D^R_G \over k-g} -{\D^R_H \over k-h}\br)\ T_{het}
 + L_0^R(int)T_{het}\cr
&={2\over k-g} \bl(\D_{G/H} +{g-h\over 2(k-h)} \D_H \br)\ T_{het}
 + L_0^R(int)T_{het}\ ,\cr} }

\no
where $L_0^R(int)$ is the internal part which does not contribute to
the spacetime metric and dilaton.
Comparing \lze\ to \hete\ we see that we can obtain the exact metric and
dilaton for the heterotic superstring if we replace $k$ in the bosonic
expressions  by $2k-h$ (except for the overall factor in the line element
$ds^2$ which is the same in both cases). In the $k\to \infty$ limit these
fields tend to their bosonic counterparts, as in \BSs, in agreement with the
general arguments of \CAL.


\newsec {Concluding remarks}

We found a general method for obtaining the conformally
exact metric and dilaton fields
for any theory which can be formulated as a coset model.
Since the value of $k$ that yields $c=26$ (or $c=15$) is actually
small, our results represent substantial deviations from the
semi-classical computations. We have
applied our method to the coset models $SO(d-1,2)/SO(d-1,1)$. Our expressions
hold true also for all the models that are obtained from this coset by
appropriate analytic continuations, e.g. $SO(d,1)/SO(d)$, $SO(d+1)/SO(d)$,
etc., by simply specializing to the appropriate region of our global space. For
the cases $d=2,3,4$ we gave explicit results. We have also derived results
that apply to the general gauged WZW model with or without supersymmetry. For
any type--II superstring the perturbative 1--loop results were shown to be
also exact except for an overall factor in the metric. We have also shown that
for a heterotic superstring the conformally exact metric and dilaton fields
can be obtained by a simple shifting of $k$ in the corresponding bosonic
expressions.  Finally, we have shown that the combination $e^{\Phi} \sqrt{-G}$
is indeed $k$--independent as conjectured in earlier work - an crucial
result for a meaningful path integral formulation.

Our exact results may be verified in perturbation theory, as was the case in
$d=2$ (the 2d heterotic case has not yet been verified). However, this should
be regarded as a challenge for perturbation theory which is beset with
uncertainties over renormalization schemes. It is generally believed, but only
tentatively proved that coset models and gauged WZW models are equivalent.
This is certainly the case at the classical level, as can be seen from the
equations of motion in the axial gauge \Bf\BSf\BSs\BSt\ . Furthermore, our
Hamiltonian approach should leave no doubt that semi-classicaly these two
theories are equivalent. The Hamiltonian versus the path integral can be
regarded as two possible approaches to quantization which may differ in higher
orders of $\hbar$, unless one insures that they are equivalent by choosing the
correct measure of integration, as we have suggested. If only to strengthen
the relationship between these two formulations, it may be of interest to
study the perturbative approach to verify our results for the more challenging
cases in $d=3,4$. This should also be helpful in pinning down the appropriate
renormalization scheme which may be useful for other computations in these
models.

We have pointed out the screening effects of the quantum corrections in the
neigborhoods of the singularities. It is not yet clear how to use these
conformally exact results in physical applications. One needs to know how this
``classical" string vacuum configuration competes with higher genus string
loop effects in specific physical situations. It may be that, together with
the small-large duality properties pointed out in \BSt\ , one may arrive at
believable physical conclusions in some regimes even from the zero genus
exact computation presented in this paper.


\bs\bs
\centerline {\bf APPENDIX }
\bs

The reader who is familiar with the exact results for the metric and dilaton
of the $2d$ Minkowski or Euclidean black hole, obtained in \DVV,
may wonder how these can be deduced in our formalism.
This appendix serves exactly this scope.
In the $2d$ case, for the coset model $SO(2,1)/SO(1,1)$,
the Casimir operators \casimir\Hcasimirleft\ take the form

\eqn\twocas{\eqalign {
 & \Delta^L_G={1\over 4}(1+x^2)^2{\partial^2\over \partial x^\mu
\partial x_\mu}-{1\over 4}(1+x^2){\partial^2\over\partial t^2}
  + {1\over 2}(1+x^2){\partial\over \partial t}\e_{\m\n}x^{\m}
{\partial\over \partial x_{\n}} \cr
 & \Delta^L_H= -{1\over 4} {\partial^2\over \partial t^2} , } }
where we have used $a_{\mu\nu}=a\epsilon_{\mu\nu}$, and $(1-a^2){\partial\over
\partial a}={\partial\over \partial t}$ for $a=\cosh t$ or $\sinh t$.
The gauge condition \gauge\ is

\eqn\twogauge{\e_{\m\n}x^{\m} {\partial\over \partial x_{\n}} T=0\ .}

\no
For gauge invariant tachyon states of the form $T=T(b,t)$, where $b$
is defined
 \foot{We have here a sign difference with (3.2) which we used in higher
dimensions. This allows us to agree with fig.2 in a previous paper \BSt\ for
2d .}
 by $b={x^2-1\over x^2+1}$, the Casimirs reduce to the
simpler equations

\eqn\twocasimir{\eqalign{&\D^L_G\ T=\bl(-(b^2-1)\partial_b ^2 -2b\partial_b
+{1\over 2(b-1)}\ \partial_t ^2 \br) T\cr
&\D^L_HT=-{1\over 4} \partial_t ^2 T\ .\cr} }

\no
Proceeding as before we obtain the line element

\eqn\twometric{ds^2=2(k-2)\bl({db^2\over 4(b^2-1)}
-\b(b) {b-1\over b+1} dt^2\br),
\qq \b^{-1}(b)=1-{2\over k} {b-1\over b+1}\ . }

\no
For the dilaton the corresponding expression is

\eqn\twodilaton{\Phi=\ln\bl({b+1\over \sqrt{\b(b)}}\br)+\Phi_0\ . }

\no
The scalar curvature reads

\eqn\curv{R={2k\over k-2} {(k-2)b+k-4\over \bl((k-2)b+k+2\br)^2}\ .}

\no
The scalar curvature is singular at $b=-(k+2)/(k-2)$ which is exactly the
point of singularity for the function $\b(b)$. To make contact with the
results of \DVV\ one needs to reparametrize $b$ in the various patches.
For instance in the black hole region outside the horizons at $b=1$ one
can set $b=\cosh 2r$, in the naked singularity region $b=-\cosh 2r'$,
and in the inside the horizons regions $b=\cos 2r''$.
The corresponding
expressions for the Euclidean black hole follow from \twometric\ and
\twodilaton\ if we analytically continue $t\to i\th$, where $\th$ is compact,
$0 < \th < 2\pi$. The cigar ($b=\cosh 2r$) and trumpet ($b=-\cosh 2r'$)
correspond to the $SO(2,1)/SO(2)$ coset, whereas the cymbal region
($b=\cos 2r''$) to the $SO(3)/SO(2)$ coset.

\listrefs
\end